\documentclass[article, a4paper,1p]{elsarticle}
\usepackage{lipsum}
\makeatletter
\def\ps@pprintTitle{%
 \let\@oddhead\@empty
 \let\@evenhead\@empty
 \def\@oddfoot{}%
 \let\@evenfoot\@oddfoot}
\makeatother
\usepackage{graphicx}
\usepackage{amsbsy}
\usepackage[usenames, dvipsnames]{color}

\usepackage[utf8x]{inputenc}
\usepackage{epstopdf}
\usepackage{amsmath,amssymb}

\usepackage{lineno,hyperref}
\begin{document}
\title{Retrieving the Covariance Matrix of an Unknown Two-Mode Gaussian State by Means of a Reference Twin Beam}
\author[slo]{Ievgen I. Arkhipov \corref{cor}}
\ead{ievgen.arkhipov01@upol.cz}
\author[slo]{Jan Pe\v{r}ina Jr.}
\ead{jan.perina.jr@upol.cz}
\address[slo]{RCPTM, Joint Laboratory of Optics of Palack\'y University and
Institute of Physics of CAS, Palack\'y University, 17. listopadu
12, 771 46 Olomouc, Czech Republic}
\cortext[cor]{Corresponding author}
\begin{abstract}
A method for revealing the covariance matrix of an unknown
two-mode Gaussian state is given based on the interference with a
reference twin beam whose covariance matrix is known. In the
method, first- and second-order cross-correlation intensity
moments are determined varying the overall phase of the reference
twin beam.
\end{abstract}
\maketitle


\section{Introduction}
The reconstruction of a state of any quantum system belongs to the
most important tasks in quantum
physics~\cite{Agarwalbook,HelstromBook,GlauberBook}. For this
reason, homodyne detection has been suggested and experimentally
implemented for the first time in Ref.~\cite{Raymer93} for quantum
light. This was the first successful example of the so-called
quantum state tomography, that provides the full information about
the analyzed quantum state. The knowledge of the quantum state of
light is extraordinarily important, as such states are useful for
testing the postulates of quantum mechanics, showing peculiar
features of quantum states (teleportation \cite{Bouwmeester1997},
dense coding~\cite{Braunstein00,Bruss04}, etc.), as well as
applying them in metrology and other applications (cryptography
\cite{Jennewein00,Gisin2002}).

The optical homodyne tomography both in its cw and pulsed variants
has become the most advanced and also powerful method in quantum
state tomography~\cite{Lvovsky09} and, as such, it has become an
indispensable technique in the field of quantum optics. The method
is based upon overlapping an unknown state with a classical light
(coherent state) with a well-defined phase, that is called a local
oscillator. The interference pattern depending on the varying
phase of the local oscillator then allows to reconstruct the
quantum state, in detail to reconstruct its Wigner function
defined in the phase space ~\cite{Agarwalbook,Lvovsky09}.
Subsequently, moments of the field operators can be obtained and
used to fully characterize nonclassicality of the analyzed quantum
state~\cite{Shchukin05r,Miranowicz10}. However, the optical
homodyne tomography is quite experimentally involved and requires
extended experimental data sets~\cite{Lvovsky09}.

For this reason, a simplified method still relying on the coherent
local oscillator has been suggested for quantum state tomography
in \cite{Banaszek1996}, and later elaborated in
\cite{Zambra2005,Bondani2009}. In this method, one of the output
ports of a beam splitter that mixes the analyzed light with the
local oscillator is monitored in general by a photon-number
resolving detector with the varying attenuation coefficient. Also
an attempt to use the data measured under different levels of
noise for quantum state tomography has been made
\cite{Harder2014}.

In many cases, the light to be analyzed has specific properties
that allow to apply simpler tools in the determination of its
state. For example, when we analyze the properties of twin beams
generated from the vacuum state by parametric
down-conversion~\cite{Perina1991Book}, the characterization by
means of the measured photocount statistics is fully sufficient
\cite{Haderka2005a,PerinaJr2012}. This possibility originates in
the generation process that does not `prescribe' specific phases
to the individual signal and idler fields. Here, the generated
light is Gaussian without coherent components, i.e. it is
completely characterized by means of its covariance matrix. Once
the covariance matrix of a given state is obtained, all the
properties of the state are easily derived. As useful examples,
entanglement of a two-mode twin beam or local nonclassicalities of
one-mode reduced states can be mentioned~\cite{Arkhipov2016}. Even
phase-space quasidistributions of integrated intensities can be
determined \cite{PerinaJr2013a}.

Whereas the analysis of covariance matrices of Gaussian two-mode
twin beams is not difficult since only certain elements of their
covariance matrices are nonzero, the measurement of covariance
matrices belonging to a general two-mode Gaussian state is more
involved. To cope with this problem, we have developed a method
for reconstructing the normally- or symmetrically-ordered
covariance matrix of such a general two-mode Gaussian state based
on mixing the analyzed state with a reference twin beam with the
varying overall phase.

Moreover, the developed approach can be extended to multimode
Gaussian states provided that the measurement on individual mode
pairs can be performed. This is useful, e.g., for spectrally
multimode twin beams composed of $ N $ paired modes. Even in this
case, the genuine multimode entanglement of the state can be
retrieved from its multimode covariance matrix
\cite{vanLoock03,Adesso06}. A specific example of this general
approach was studied by Altepeter {\it {et
al}}.~\cite{Altepeter04} who reconstructed the polarization state
of a photon pair by acquiring photon coincidence-count statistics.

The paper is organized as follows. Description of general two-mode
Gaussian states through their covariance matrices is given in
Sec.~II. Twin beams as the most common kind of two-mode Gaussian
states are discussed in Sec.~III. The method for revealing the
covariance matrix of a general two-mode Gaussian state is given in
Sec.~IV. Sec.~V brings conclusions.

\section{General two-mode Gaussian states}

The normally-ordered characteristic function $ C_{\mathcal N}$ for
a general two-mode Gaussian state is defined as follows
\begin{equation}\label{CF1}    
C_{\mathcal N}(\beta_{1},\beta_{2}) =
\mathrm{Tr}\left[\hat\rho(0)\exp (\beta_{1}\hat
a^{\dagger}_{1}+\beta_{2}\hat
a^{\dagger}_{2})\exp(-\beta^{\ast}_{1}\hat a_{1}
-\beta^{\ast}_{2}\hat a_{2})\right],
\end{equation}
where Tr stands for the operator trace and $\hat a_i$ ($\hat
a_i^{\dagger}$) denote the boson annihilation (creation) operators
of mode $i$, $ i=1,2 $.

The normally-ordered characteristic function $C_{\mathcal N} $ in
Eq.~(\ref{CF1}) can be expressed via its complex covariance matrix
$ {\bf A}_{\cal N} $,
\begin{eqnarray} \label{CM}  
 & {\bf A}_{\cal N} = \left[ \begin{array}{cc} {\bf B}_1 & {\bf
  D}_{12}  \cr {\bf D}_{12}^\dagger & {\bf B}_2 \end{array} \right]
  ,& \\
 &\hspace{5mm} {\bf B}_j = \left[ \begin{array}{cc} -B_j & C_j
   \cr C_j^* & -B_j \end{array} \right] , \hspace{3mm} j=1,2,
  \hspace{3mm} {\bf D}_{12} = \left[ \begin{array}{cc} \bar{D}_{12}^* & {D}_{12}
  \cr {D}_{12}^* & \bar{D}_{12} \end{array} \right], &
\end{eqnarray}
in the form $ C_{\cal N}(\hat{\mbox{\boldmath$\beta$}}) =\exp (
\hat{\mbox{\boldmath$ \beta $}}^\dagger {\bf A}_{\mathcal N}
\hat{\mbox{\boldmath$ \beta $}}/2 ) $, where $
\hat{\mbox{\boldmath$ \beta $}} =
(\beta_1,\beta_1^{\ast},\beta_2,\beta_2^{\ast})^T$. The
coefficients $ B_j $, $ C_j $, $ j=1,2 $, $ \bar D_{12} $ and $
D_{12} $ occurring in Eq.~(\ref{CM}) are defined as
\begin{eqnarray} \label{qnc}   
B_j=\langle\Delta\hat a^{\dagger}_{j}\Delta\hat a_{j}\rangle, & & C_j=\langle\Delta\hat a_{j}^2\rangle, \nonumber \\
D_{12} =\langle\Delta\hat a_{1}\Delta\hat a_{2}\rangle, & &  \bar
D_{12} =-\langle\Delta\hat a_{1}^{\dagger}\Delta\hat a_{2}\rangle
.
\end{eqnarray}
The determination of coefficients $ B_j $ and $ C_j $, $ j=1,2 $,
together with the coefficients $ D_{12} $ and $ \bar{D}_{12} $ is
thus sufficient to fully characterize a general two-mode Gaussian
field without coherent components.  All possible correlation
functions can then be easily derived \cite{Perina1991Book}.

The coefficients occurring in the normally-ordered covariance
matrix $\bf A_{\mathcal N}$ give also the coefficients in the
symmetrically-ordered covariance matrix $\bf A_{\mathcal
S}$~\cite{Arkhipov2016a}, which is important for the determination
of entanglement. Indeed, the determinant $I_4 \equiv {\rm det}
(\bf A_{\mathcal S}) $ of the symmetrically-ordered covariance
matrix along with the local unitary invariants $ I_j $, $ j=1,2,3,
$ and $ I_{j\mathcal S} $, $ j=1,2 $ of the normally- and
symmetrically-ordered covariance matrices, respectively,
completely quantify both nonclassicality and entanglement of the
state~(for details, see \cite{Arkhipov2016}).

\section{Pure twin beams}

To determine the needed coefficients, we use as a reference a pure
twin beam with a specific form of its covariance matrix. Pure twin
beams represent a specific form of two-mode Gaussian states that
is standardly generated in the process of parametric
down-conversion. The boson operators characterizing the emitted
signal ($ \hat a_1^{\rm out} $) and idler ($ \hat a_2^{\rm out} $)
fields are written in the Heisenberg picture as
follows~\cite{Perina1991Book}:
\begin{eqnarray}\label{heiseq}
\hat a_1^{\rm out}  &=& \cosh(\sqrt{G})\hat a_1^{\rm in} +i\exp(i\phi)\sinh(\sqrt{G})\hat a_2^{{\rm in}\dagger}, \nonumber \\
\hat a_2^{\rm out} &=& \cosh(\sqrt{G})\hat a_2^{\rm
in}+i\exp(i\phi)\sinh(\sqrt{G})\hat a_1^{{\rm in}\dagger},
\end{eqnarray}
where $G$ is the gain of the parametric process, $ \hat a_1^{\rm
in} $ ($ \hat a_2^{\rm in} $) denotes the incident signal-
(idler-) field annihilation operator and $ \phi $ stays for a
phase that follows the phase of the coherent pump field.

Assuming the incident vacuum state in both the signal and idler
fields, we arrive at the following only nonzero coefficients of
the normally-ordered covariance matrix of this reference beam:
\begin{equation}\label{twb}
 B_{1,2}^{R} = B_p,  \quad
 D_{12}^{R}=i\exp(i\phi)\sqrt{B_p(B_p+1)}
\end{equation}
and $B_{\rm p}=\sinh^2(\sqrt{G})$ is the mean photon-pair number.

The problem how to reconstruct the coefficients of the
normally-ordered covariance matrix ${\bf A}_{\cal N}^{R}$ of the
reference twin beam has been discussed in detail in
Refs.~\cite{PerinaKrepelka11,Allevi2013,PerinaJr2012}.

\section{Retrieving the covariance matrix of an unknown two-mode Guaussian state}

The scheme for retrieving the covariance matrix of an unknown
Gaussian state with vanishing coherent components is shown in
Fig.~1.
\begin{figure}
\center
\includegraphics[width=0.4\textwidth]{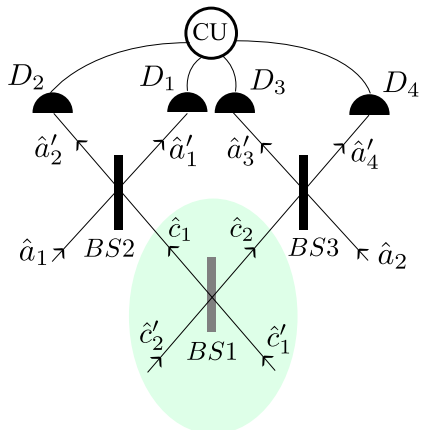}
\caption{The experimental scheme. Two modes ($\hat c'_1$ and $\hat
c'_2$) of a pure twin beam are mixed on beam splitter BS1 to
provide a reference two-mode field ($\hat c_1$ and $\hat c_2$).
The output modes $\hat c_1$ and $\hat c_2$ of beam splitter BS1
are combined with two modes $\hat a_1$ and $\hat a_2$ of an
unknown two-mode Gaussian state at balanced beam splitters BS2 and
BS3. The photocount statistics of four output modes $ \hat a'_j $,
$ j=1,\ldots,4 $, leaving beam splitters BS2 and BS3 are measured
by detectors $D_1$, $D_2$, $D_3$, $D_4$ and correlation unit CU.}
\end{figure}
It relies on mixing the analyzed state with a reference twin beam.
However, a pure twin beam composed of only photon pairs and
exhibiting the thermal photon-number statistics in the signal and
idler fields is not sufficient for this task that requires all the
coefficients of the reference covariance matrix being nonzero. For
this reason, we first mix the signal (annihilation operator $
\hat{c}'_1 $) and idler ($ \hat{c}'_2 $) fields on a beam splitter
BS1 with the varying transmissivity $ t_1 $. At the output ports
of beam splitter BS1 and depending on the transmissivity $ t_1 $,
there occur different kinds of states useful in the reconstruction
\cite{Arkhipov2016,Arkhipov2016a}. In the proposed method, the
reference light at the output ports ($ \hat{c}_1 $ and $ \hat{c}_2
$) of beams splitter BS1 is superimposed with the analyzed
two-mode Gaussian state at balanced beam splitters BS2 and BS3.
The output ports ($ \hat{a'}_j $, $ j=1,\ldots,4 $) of beam
splitters BS2 and BS3 are then monitored by four detectors
measured in coincidence.

The unitary transformations describing the functioning of three
beam splitters BS${}_j$ with amplitude transmissivities $ t_j $
and phase shifts $ \theta_j $, $ j=1,2,3 $, are expressed in
general as follows:
\begin{eqnarray}\label{IO}
\begin{pmatrix}
\hat c_1 \\
\hat c_2
\end{pmatrix}&=&
\begin{pmatrix}
t_1 & r_1\exp(i\theta_1) \\
-r_1\exp(-i\theta_1) & t_1
\end{pmatrix}
\begin{pmatrix}
\hat c'_1 \\
\hat c'_2
\end{pmatrix}, \nonumber \\
\begin{pmatrix}
\hat a'_1 \\
\hat a'_2
\end{pmatrix} &=&
\begin{pmatrix}
t_2 & r_2\exp(i\theta_2) \\
-r_2\exp(-i\theta_2) & t_2
\end{pmatrix}
\begin{pmatrix}
\hat a_1 \\
\hat c_1
\end{pmatrix},
\nonumber \\ \nonumber \\
\begin{pmatrix}
\hat a'_3 \\
\hat a'_4
\end{pmatrix} &=&
\begin{pmatrix}
t_3 & r_2\exp(i\theta_3) \\
-r_3\exp(-i\theta_3) & t_3
\end{pmatrix}
\begin{pmatrix}
\hat a_2 \\
\hat c_2
\end{pmatrix},
\end{eqnarray}
where the annihilation operators $ \hat{a}_1 $ and $ \hat{a}_2 $
belong to the modes of the analyzed two-mode Gaussian state.

Assuming the balanced beam splitters BS2 and BS3 ($
t_2=t_3=1/\sqrt{2} $) with zero phase shifts ($ \theta_2=\theta_3
=0 $) and applying the relations in Eqs.~(\ref{IO}), we reveal the
following formulas giving the number operators $ \hat{n}'_j $ of
fields at the detectors as functions of the operators of the
analyzed state and the reference state:
\begin{eqnarray}\label{av_n}
 \hat n'_1 &=& \hat {a'}_{1}^{\dagger}\hat a'_1 = \frac{1}{2}\left(\hat a_1^{\dagger}\hat a_1+
  \hat a_1^{\dagger}\hat c_1+\hat a_1\hat c_1^{\dagger}+\hat c_1^{\dagger}\hat c_1\right), \nonumber \\
 \hat n'_2 &=& \hat {a'}_{2}^{\dagger}\hat a'_2 = \frac{1}{2}\left(\hat a_1^{\dagger}\hat a_1-
  \hat a_1^{\dagger}\hat c_1-\hat a_1\hat c_1^{\dagger}+\hat c_1^{\dagger}\hat c_1\right), \nonumber \\
 \hat n'_3 &=& \hat {a'}_{3}^{\dagger}\hat a'_3 = \frac{1}{2}\left(\hat a_2^{\dagger}\hat a_2+
  \hat a_2^{\dagger}\hat c_2+\hat a_2\hat c_2^{\dagger}+\hat c_2^{\dagger}\hat c_2\right), \nonumber \\
 \hat n'_4 &=& \hat {a'}_{4}^{\dagger}\hat a'_4 =
  \frac{1}{2}\left(\hat a_2^{\dagger}\hat a_2-\hat a_2^{\dagger}\hat
  c_2-\hat a_2\hat c_2^{\dagger}+\hat c_2^{\dagger}\hat c_2\right).
\end{eqnarray}
The normally-ordered characteristic function $ C_{\cal N} $ of the
four-mode Gaussian state characterizing the four fields in front
of detectors is written as
\begin{eqnarray}\label{CF4_1}
C_{\cal N}(\beta_1,\beta_2,\beta_3,\beta_4)=\mathrm{Tr}\left[\hat\rho'(0)\exp \left(\sum\limits_{i=1}^4\beta_{i}\hat
{a'}^{\dagger}_{i}\right)\exp\left(-\sum\limits_{i=1}^4\beta^{\ast}_{i}\hat a'_{i}\right)\right].
\end{eqnarray}
The quantum-mechanical averaging in Eq.~(\ref{CF4_1}) is performed
by the statistical operator
$\hat\rho'(0)=\hat\rho_{12}(0)\otimes\hat\rho^R(0)$, where
$\hat\rho_{12}$ is the statistical operator of the unknown
two-mode Gaussian state and the operator $\hat\rho^R$ describes
the incident reference twin beam.

Given Eqs.~(\ref{qnc}) and~(\ref{CF4_1}), the normally-ordered
characteristic function $ C_{\cal N} $ is obtained in the form:
\begin{eqnarray}\label{CF4}
 && \hspace{-7mm} C_{\cal N}(\beta_1,\beta_2,\beta_3,\beta_4)= \nonumber \\
 && \hspace{-2mm} \exp\left\{-\sum\limits_{i=1}^4B'_i|\beta_i|^2+
  \left[\frac{1}{2}\sum\limits_{i=1}^4C'_i\beta_i^{\ast2}+
  \sum\limits_{j<k}^4D'_{jk}\beta_j^{\ast}\beta_k^{\ast}+\sum\limits_{j<k}^4\bar
  D'_{jk}\beta_j\beta_k^{\ast}+{\rm c.c.}\right]\right\},
   \nonumber \\
 &&
\end{eqnarray}
where the coefficients $ B'_i $, $ C'_i $, $ D'_{jk} $, and $
\bar{D}'_{jk} $ are determined by the formulas written in
Eqs.~(\ref{qnc}).

Denoting the normally-ordered photon-number moments $\langle\hat
n^k\rangle_{\cal N}$ as moments $ \langle W^k\rangle_{\cal N}$ of
the integrated intensities, as suggested by the photodetection
theory \cite{Perina1991Book}, we derive the second-order
correlations of the integrated-intensity fluctuations $ \Delta W $
in different modes from Eq.~(\ref{CF4}) in the form:
\begin{eqnarray}\label{WjWk}
 && \hspace{-7mm} \langle\Delta W_j\Delta W_k\rangle_{\cal N} =\langle\hat {a'}_j^{\dagger}\hat {a'}_k^{\dagger}\hat a'_j
  \hat a'_k\rangle-\langle\hat {a'}_j^{\dagger}\hat a'_j\rangle\langle\hat {a'}_k^{\dagger}
  \hat a'_k\rangle \nonumber \\
 && \hspace{-5mm} = \left.\frac{\partial^4C_{\cal N}}{\partial\beta_j
  \partial(-\beta_j^{*})\partial\beta_k\partial(-\beta_k^*)}\right|_{\{\beta_j\}=\{\beta_j^*\}=0}-
  \left.\frac{\partial^2C_{\cal N}}{\partial\beta_j\partial(-\beta_j^{*})}
  \frac{\partial^2C_{\cal N}}{\partial\beta_k\partial(-\beta_k^*)}\right|_{\{\beta_j\}=\{\beta_j^*\}=0} \nonumber \\
&& \hspace{-5mm} =|D'_{jk}|^2+|\bar D'_{jk}|^2,
 \hspace{5mm} j\neq k.
\end{eqnarray}
Now, applying the photodetection theory for the detectors with
quantum detection efficiencies $ \eta_j $ and dark-count rates $
n_{dj} $, we arrive at the following second-order moments of
photocount fluctuations at all four detectors:
\begin{equation}\label{correl1}
 \langle\Delta\hat m_j\Delta\hat m_k \rangle = \eta_j\eta_k\langle\Delta W_j\Delta W_k\rangle_{\cal N},
  \quad j \neq k.
\end{equation}
Applying further Eqs.~(\ref{qnc}),~(\ref{IO}) and (\ref{WjWk}), we
reveal the following second-order moments of photocount
fluctuations:
\begin{eqnarray}\label{correl2}
\langle\Delta\hat m_1\Delta\hat m_2\rangle &=& \frac{\eta_1\eta_2}{4}\Big(B_1^2+|C_1|^2+B_{1}^{R2}+|C_1^R|^2-2B_1B_{1}^{R}-2{\rm Re}\{C_1C_1^{R\ast}\}\Big), \nonumber \\
\langle\Delta\hat m_3\Delta\hat m_4\rangle &=& \frac{\eta_3\eta_4}{4}\Big(B_2^2+|C_2|^2+B_{2}^{R2}+|C_2^R|^2-2B_2B_{2}^{R}-2{\rm Re}\{C_2C_2^{R\ast}\}\Big), \nonumber \\
\langle\Delta\hat m_1\Delta\hat m_3\rangle &=& \frac{\eta_1\eta_3}{4}\Big(|D_{12}|^2+|\bar D_{12}|^2+|D_{12}^R|^2+|\bar D_{12}^R|^2+2{\rm Re}\{D_{12}D_{12}^{R\ast}\}+ \nonumber \\
&&2{\rm Re}\{\bar D_{12}\bar D_{12}^{R\ast}\}\Big),\nonumber \\
\langle\Delta\hat m_1\Delta\hat m_4\rangle &=& \frac{\eta_1\eta_4}{4}\Big(|D_{12}|^2+|\bar D_{12}|^2+|D_{12}^R|^2+|\bar D_{12}^R|^2-2{\rm Re}\{D_{12}D_{12}^{R\ast}\}- \nonumber \\
&&2{\rm Re}\{\bar D_{12}\bar D_{12}^{R\ast}\}\Big), \nonumber \\
\langle\Delta\hat m_2\Delta\hat m_3\rangle &=& \frac{\eta_2\eta_3}{4}\Big(|D_{12}|^2+|\bar D_{12}|^2+|D_{12}^R|^2+|\bar D_{12}^R|^2-2{\rm Re}\{D_{12}D_{12}^{R\ast}\}- \nonumber \\
&&2{\rm Re}\{\bar D_{12}\bar D_{12}^{R\ast}\}\Big), \nonumber \\
\langle\Delta\hat m_2\Delta\hat m_4\rangle &=& \frac{\eta_2\eta_4}{4}\Big(|D_{12}|^2+|\bar D_{12}|^2+|D_{12}^R|^2+|\bar D_{12}^R|^2+2{\rm Re}\{D_{12}D_{12}^{R\ast}\}+ \nonumber \\
&&2{\rm Re}\{\bar D_{12}\bar D_{12}^{R\ast}\}\Big).
\end{eqnarray}
The formulas in Eqs.~(\ref{correl2}), when applied to the analyzed
two-mode Gaussian state and the reference twin beam, allow to
recover all coefficients of the covariance matrix of the analyzed
state. The determination of the coefficients is naturally split
into the following four steps.

\textit{Retrieving the coefficients $B_1$ and $B_2$} --- These
coefficients give the mean numbers of photons present in both
modes of the analyzed state. If the inputs of the reference field
are replaced by the vacuum, we immediately arrive at the values of
these coefficients using the relations in Eqs.~(\ref{av_n}):
\begin{eqnarray}\label{correl3}  
 B_1 = 2 \left( \langle \hat m_1 \rangle - n_{d1} \right) /
 \eta_1, & &
 B_2 = 2 \left( \langle \hat m_3 \rangle - n_{d3} \right) /
 \eta_3.
\end{eqnarray}

\textit{Retrieving the coefficients $C_1$ and $C_2$} --- To reveal
these coefficients, we exploit the fact that a pure twin beam with
the mean photon-pair number $ B_p $ gives two separable squeezed
states with opposite phases $ \phi $ when its constituents are
combined at the balanced beam splitter BS1~\cite{Paris97}. Thus,
the reference field attains the coefficients $ B_1^R = B_2^R =
B_{\rm p} $, $ C_{1}^R=-C_{2}^R=i\exp(i\phi)\sqrt{B_{\rm p}(B_{\rm
p}+1)} $~\cite{Arkhipov2016a} and the suitable relations in
Eqs.~(\ref{correl2}) can be recast into the form:
\begin{eqnarray}
 \langle \Delta \hat m_1\Delta\hat m_2\rangle &=& \frac{\eta_1\eta_2}{4}
  \Big(2B_{\rm p}^2+B_{\rm p}(1-2B_1)+  \langle\Delta W_1^2\rangle_{\cal N} \Big)
  - \nonumber \\
&& \frac{\eta_1\eta_2\sqrt{B_{\rm p}(B_{\rm p}+1)}}{2} {\rm Im}\left\{ \exp\left(-i\phi\right)C_{1} \right\} \label{c1c21},\nonumber \\
 \langle\Delta\hat m_3\Delta\hat m_4\rangle &=&
  \frac{\eta_3\eta_4}{4}\Big(2B_{\rm p}^2+B_{\rm p}(1-2B_2)+ \langle\Delta W_2^2\rangle_{\cal N} \Big) + \nonumber \\
&&\frac{\eta_3\eta_4\sqrt{B_{\rm p}(B_{\rm
   p}+1)}}{2} {\rm Im} \left\{ \exp\left(-i\phi\right)C_{2} \right\} \label{c1c22}.
\end{eqnarray}
The formulas in Eq.~(\ref{c1c22}) allow us to determine the
variances $ \langle\Delta W_j^2\rangle_{\cal N} $ for $ j=1,2 $ of
the constituents of the analyzed field provided that the reference
field is absent. The usual formula for the second moment $ \langle
W_j^2\rangle_{\cal N} $ of integrated intensity of mode $ j $, $
\langle W_j^2\rangle_{\cal N} = 2B_j^2+|C_j|^2 $, can be recast
into that for the variance $ \langle\Delta W_j^2\rangle_{\cal N}
=B_j^2+|C_j|^2 $, that immediately provides the absolute value $
|C_j| $. For the complex coefficients $ C_j $, we need to vary the
phase $ \phi $ of the reference field, that is derived from the
pump field that created the reference pure twin beam. The obtained
interference pattern then gives us both the magnitudes and phases
of both coefficients.

If the analyzed state is known to be symmetric ($B_1=B_2$ and
$C_1=C_2\equiv C$), we can even apply the following simpler
formula to arrive at the coefficient $ C $:
\begin{eqnarray}
\frac{1}{\sqrt{B_{\rm p}(B_{\rm p}+1)}}\left(\frac{\langle\Delta\hat m_3\Delta\hat m_4\rangle}{\eta_3\eta_4}-\frac{\langle\Delta\hat
m_1\Delta\hat m_2\rangle}{\eta_1\eta_2}\right)=
{\rm Im} \left\{ \exp\left(-i\phi\right)C \right\}.
\end{eqnarray}

{\textit{Retrieving the coefficient $ D_{12}$}} --- We need as a
reference field the original pure twin beam for which $ D_{12}^R $
is given in Eq.~(\ref{twb}) and $ \bar D_{12}^R $ vanishes. The
third and fourth relations in Eqs.~(\ref{correl2})  can be rearranged into the formula:
\begin{equation}\label{d12}
\frac{1}{\sqrt{B_{\rm p}(B_{\rm p}+1)}}\left(\frac{\langle\Delta\hat m_1\Delta\hat m_3\rangle}{\eta_1\eta_3}-\frac{\langle\Delta\hat m_1\Delta\hat
m_4\rangle}{\eta_1\eta_4}\right)= {\rm Im} \left\{
\exp\left(-i\phi\right) D_{12}\right\}.
\end{equation}
According to Eq.~(\ref{d12}), the variation of the pump phase
$\phi$ provided both the magnitude and phase of coefficient $
D_{12} $. We note that also other combinations of the second-order
moments in Eqs.~(\ref{correl2}) can be used to reveal the
coefficient $ D_{12} $.

\textit{Retrieving the coefficient $\bar D_{12}$} --- To retrieve
the coefficient $\bar D_{12}$ one needs a nonzero coefficient
$\bar D_{12}^R$ of the reference field. Such coefficient cannot be
obtained by a simple mixing of the constituents of a pure twin
beam on beam splitter BS1. However, if we consider only one
constituent of the pure twin beam and mix it with the vacuum state
of beam splitter BS1 with transmissivity $ t_1 = 1/\sqrt{2} $, we
arrive at the fields with zero values $ D_{12}^R$, $C_1^R $ and
$C_2^R $, but nonzero coefficients $ B_1^R = B_2^R = B_{\rm p}/2 $
and $\bar D_{12}^R=\pm B_{\rm p}/2$, where the plus (minus) sign
is taken for mode $\hat c'_2$ ($\hat c'_1$) in the vacuum state.
In this case, the following relation is revealed:
\begin{equation}\label{bd12}
 \frac{\langle\Delta\hat m_1\Delta\hat m_3\rangle}{ \eta_1\eta_3} - \frac{\langle
  \Delta\hat m_1\Delta\hat m_4\rangle}{\eta_1\eta_4} = {\rm Re} \{ \bar D_{12}^R \bar
  D_{12}^* \} .
\end{equation}
The formula in Eq.~(\ref{bd12}) suggests that the variation of
complex phase of the reference coefficient $ \bar D_{12}^R $
allows to recover the coefficient $ \bar D_{12} $ of the analyzed
field. This can easily be accomplished by imposing a variable
phase shift $\theta $ to, e.g., mode $\hat c_1$ by a phase
modulator placed between the beam splitters BS1 and BS2 [$\hat c_1
\rightarrow \exp(i\theta)\hat c_1$]. In this case,
Eq.~(\ref{bd12}) is transformed into the form
\begin{eqnarray}
  \frac{2}{B_{\rm p}}\left(\frac{\langle\Delta\hat m_1\Delta\hat m_3\rangle}{ \eta_1\eta_3} - \frac{\langle
  \Delta\hat m_1\Delta\hat m_4\rangle}{\eta_1\eta_4}\right) = \pm{\rm Re} \{ \exp(-i\theta)  \bar
  D_{12}^* \},
\label{22}
\end{eqnarray}
where again the plus (minus) sign is taken for the mode $\hat
c'_2$ ($\hat c'_1$) in the vacuum state. According to
Eq.~(\ref{22}) the variation of phase $\theta $ then provides both
the real and imaginary part of the coefficient $\bar D_{12}$.

In the experiment, a source of identical reference pure twin beams
is needed. If we consider such two beams in the analyzed scheme,
one as a reference beam and the other as a beam in an unknown
state, the parameters of the reference twin beam can be reached.
This allows to check the quality of the applied reference twin
beam, that is assumed to be ideally composed only of photon pairs.

At the end, we note that the developed method can be generalized
to allow for the characterization of two-mode Gaussian states with
nonzero coherent components. In this case, the coherent components
in both modes of the analyzed state have to be identified first,
by applying the homodyne detection scheme. Then, the above written
formulas can be generalized to include the coherent components. So
the contributions from coherent components can easily be
subtracted.

\section{Conclusions}

We have suggested a method for characterizing a general two-mode
Gaussian state with vanishing coherent components. The
coefficients of its normally-ordered covariance matrix are
revealed by mixing the analyzed state with a reference beam
obtained from a pure twin beam, by using two balanced beam
splitters. The variation of the phase of the pump beam that
generates the reference twin beam together with the variation of
the phase of one mode of the reference beam are needed in the
method that monitors the first- and second-order moments of
photocounts at four detectors placed in the experimental setup.

\section*{Acknowledgments}
The authors thank J. Pe\v{r}ina and O. Haderka for discussions.
This work was supported by the projects No.~15-08971S of the GA
\v{C}R and No.~LO1305 of the M\v{S}MT \v{C}R. I.A. thanks
project IGA\_PrF\_2016\_002 of IGA UP Olomouc.

\end{document}